\documentclass[twocolumn,fp]{jpsj3} 
\usepackage{txfonts}
\usepackage{times}
\usepackage{color}
\usepackage{amsmath,amsbsy,amssymb,graphics}
\usepackage{bm}
\usepackage{tabularx}
\newcommand{\p}{\partial}

\newcommand{\hd}{\mathcal{H}}
\newcommand{\gd}{\mathcal{G}}

\title{
Anisotropic Magnetic Response in Kondo Lattice with Antiferromagnetic Order
}

\author{Taku Kikuchi$^1$, Shintaro Hoshino$^2$ and Yoshio Kuramoto$^1$}

\inst{
$^1$Department of Physics, Tohoku University, Sendai 980-8578, Japan
\\
$^2$Department of Basic Science, The University of Tokyo, Meguro 153-8902, Japan
}

\recdate{\today}

\abst{
Magnetic properties are investigated for the Kondo lattice by using the continuous time quantum Monte Carlo (CT-QMC) and the dynamical mean field theory (DMFT).
The DMFT+CT-QMC approach is extended so as to derive the anisotropic magnetic response in the antiferromagnetic phase.
The longitudinal and transverse magnetic susceptibilities are numerically calculated in the antiferromagnetic phase.
For the RKKY regime with a small Kondo coupling, the transverse susceptibility does not decrease below the transition temperature while the longitudinal susceptibility decreases as expected from the mean field picture.
In the competing region between the RKKY interaction and the Kondo effect, however, both longitudinal and transverse susceptibilities decrease below the transition temperature.
The obtained results naturally explain the temperature dependence of the magnetic susceptibility observed in CeT$_2$Al$_{10}$ ($T$=Ru,Os,Fe) family.
}

\kword{
CeT$_2$Al$_{10}$,
continuous-time quantum Monte Carlo,
dynamical mean-field theory,
magnetic susceptibility,
two-particle Green function
}

\begin{document}
\maketitle

\section{Introduction}

Heavy electron systems show various intriguing phenomena caused by interactions between nearly localized $f$ electrons and conduction ($c$) electrons.
Recently the Kondo-insulator family CeT$_2$Al$_{10}$ (T=Ru,Os,Fe) \cite{nishioka2009novel,muro2010structural} attract much attention due to their unusual magnetic properties.
CeRu$_2$Al$_{10}$ and CeOs$_2$Al$_{10}$ show antiferromagnetic (AFM) orders at low temperatures \cite{robert2012anisotropic,khalyavin2010long,mignot2011neutron,kato2011magnetic,adroja2010long}.
The magnetic susceptibility starts to decrease for all the directions slightly above the AFM transition temperature \cite{nishioka2009novel,muro2010structural,takesaka2010semiconducting,kondo2011high,tanida2010existence}.
This decreasing behavior continues even inside the AFM phase.
According to the mean-field theory, only the longitudinal susceptibility decreases below the magnetic transition temperature.
On the other hand, CeFe$_2$Al$_{10}$, which remains paramagnetic down to the lowest experimentally accessible temperature, shows a typical Kondo-insulator behavior.
For example, the susceptibility in this compound has a peaked structure around 80K \cite{muro2010semiconducting}.
A motivation of this paper is to understand these magnetic susceptibilities of CeT$_2$Al$_{10}$,
 which might originate from the interplay between the RKKY interaction and the Kondo effect.

The simplest description for Kondo insulators is given by the Kondo lattice model with one $c$ electron per site (half filling).
In this model, the periodically aligned localized $f$ electrons are coupled to $c$ electrons by the Kondo exchange interaction $J$ at each site.
For small Kondo interactions, the system has a magnetic ground state by the RKKY interaction.
If the Kondo energy dominates the RKKY one at larger $J$, on the other hand, the paramagnetic Kondo insulator is realized.
Thus we have the quantum critical point at $J=J_c$ where the magnetic state changes into paramagnetic one in the ground state,
 which is included in the Doniach phase diagram \cite{doniach1977kondo}.
This quantum phase transition has been actively investigated \cite{lacroix1979phase, fazekas1991magnetic, assaad1999quantum, peters2007magnetic, otsuki2009kondo, hoshino2010itinerant}.

For analysis of the Kondo lattice model, a framework using the dynamical mean-field theory (DMFT) combined with continuous-time quantum Monte Carlo (CT-QMC),
 has reasonably reproduced the finite-temperature Doniach phase diagram \cite{otsuki2009kondo}.
By using this framework, we numerically derive the anisotropic magnetic properties in this paper.
For this purpose, we extend the previous $J$-expansion CT-QMC method \cite{otsuki2007continuous} to the one that can also treat transverse magnetization.
This makes it possible to calculate transverse moments and transverse magnetic susceptibilities.
Especially we focus on the behavior of susceptibilities in the competing region between the RKKY interaction and the Kondo effect.

The most direct way to estimate the susceptibility is to apply a small magnetic field and measure the magnetic moment.
We can alternatively use two-particle Green functions for evaluation of susceptibilities.
With use of this method, we can decompose the susceptibility into each contribution from $c$ and $f$ electrons.
The framework has been formulated for longitudinal susceptibilities \cite{otsuki2008kondo}.
In this paper we describe the calculation method for transverse susceptibilities.

This paper is organized as follows.
In the next section, we introduce the Kondo lattice model and extend the DMFT and $J$-expansion CT-QMC to the systems with transverse magnetizations.
Section 3 provides numerical results for magnetic-field and Kondo interaction dependences of magnetization.
We discuss in \S4 magnetic susceptibilities, and make a comparison with experimental results in CeT$_2$Al$_{10}$.
We summarize the results in \S5.
The Appendix describes how to derive susceptibilities from two-particle Green functions.

\section{Model and Extension of DMFT+CTQMC method}

\subsection{Kondo lattice model}

In this paper, we deal with the bipartite Kondo lattice model in magnetic fields.
The Hamiltonian with hopping only between different sublattices is given by
\begin{align}
\hd_{\text{KL}} &= \sum_{\bm{k}}' \sum_\sigma \epsilon_{\bm{k}} \left( c_{\bm{k}\text{A}\sigma}^{\dagger} c_{\bm{k}\text{B}\sigma} + \text{h.c.} \right)
 - \sum_{\lambda} \sum_{\bm{k}}' \sum_{\sigma} \mu c_{\bm{k}\lambda\sigma}^{\dagger} c_{\bm{k}\lambda\sigma} \notag \\
 &+ \sum_{\lambda} \sum_{i\in \{ \lambda \} } 2J\bm{S}_{f,i}^{\lambda} \cdot \bm{S}_{c,i}^{\lambda} 
- \sum_{\lambda} \sum_{i\in \{\lambda \} }  H_z (\bm{S}_{f,i}^{\lambda} + \bm{S}_{c,i}^{\lambda}) , 
\label{eq:1-1}
\end{align}
where $\lambda = {\rm A, B}$ and $\sigma = \uparrow,\downarrow$ are the sublattice and spin indices, respectively.
The summation $\sum_{\bm k}'$ is taken over the half Brillouin zone, and $\sum_{i\in \{ \lambda \}}$ over the sites which belong to the sublattice $\lambda$.
$\bm{S}_{c,i}^{\lambda} = (1/2)\sum_{\sigma\sigma'} c_{i\lambda\sigma}^{\dagger} \bm{\sigma}_{\sigma\sigma'} c_{i\lambda\sigma'}$ and $\bm{S}_{f,i}^{\lambda} = (1/2)\sum_{\sigma\sigma'} \bm{\sigma}_{\sigma\sigma'} X^\lambda_{i\sigma\sigma'}$ are $c$ and $f$ electron spins at site $i$ belonging to  the sublattice $\lambda$.
Here $c_{i\lambda\sigma}^{\dagger}$ and $c_{i\lambda\sigma}$ are the creation and annihilation operators of $c$ electrons, and the operator $X^\lambda_{i\sigma\sigma'}$ changes the localized-spin state from $\sigma'$ to $\sigma$.
The fourth term is the Zeeman energy, where the g-factor for $c$ electrons is assumed to be the same as the one for localized spins.
Although we take the chemical potential $\mu=0$ in numerical calculations, we consider a general case in the formalism.

The density of states for $c$ electrons is chosen as the semi-circular shape $\rho_0 (\epsilon) = (2/\pi D) \times \sqrt{1-(\epsilon/D)^2}$.
We take $D = 1$ as a unit of energy.
As stated in Introduction, we apply the DMFT\cite{georges1996dynamical,otsuki2008kondo} to the Kondo lattice in this paper, and the CT-QMC \cite{gull2011continuous,otsuki2007continuous} for the effective impurity problem of the DMFT.
In the next two subsections, we explain the extension of this scheme to the case with transverse magnetic moments.

\subsection{Local self-energy in DMFT}

With a transverse magnetization, the self-energy has off-diagonal components with respect to spin index.
Then we need to consider the following $2\times 2$ self-energy matrix :
\begin{equation}	
\hat{\Sigma }^\lambda (i \epsilon_n) = 
\begin{pmatrix}
\Sigma_{\uparrow \uparrow}^\lambda (i \epsilon_n) & \Sigma_{\uparrow \downarrow}^\lambda (i \epsilon_n) \\
\Sigma_{\downarrow \uparrow}^\lambda (i \epsilon_n) & \Sigma_{\downarrow \downarrow}^\lambda (i \epsilon_n)
\end{pmatrix} 
,
\label{eq:2-1}
\end{equation}		
where $\epsilon_n = (2n+1)\pi/\beta$.
Note that the self-energies are independent of wavenumber in the DMFT, but depend on the sublattice.
With A and B sublattices combined, the Green function of $c$ electrons is given by the $4\times 4$ matrix as
\begin{align}	
&\hat{\bar {\bm G}}_{c,\bm k}  (i \epsilon_n) =
\left(
 \begin{array}{cc}
 \hat{S}^{\text{A}} & -\hat{E} \\
 -\hat{E} & \hat{S}^{\text{B}} \\
\end{array}
\right)^{-1}, \notag \\
&\hat{S}^{\lambda} = \left(
 \begin{array}{cc}
 i \epsilon_n + \mu - H_z - \Sigma_{\uparrow \uparrow}^{\lambda}(i \epsilon_n) & -\Sigma_{\uparrow \downarrow}^{\lambda}(i \epsilon_n)  \\
 -\Sigma_{ \downarrow\uparrow}^{\lambda}(i \epsilon_n) & i \epsilon_n + \mu + H_z - \Sigma_{\downarrow \downarrow}^{\lambda} (i \epsilon_n) \\
\end{array}
\right), \notag \\
&\hat{E} = \left(
 \begin{array}{cc}
  \epsilon_{\bm{k}} & 0 \\
  0 & \epsilon_{\bm{k}} \\
\end{array}
\right).
\label{eq:2-3}
\end{align}		
We define the local Green function by
\begin{align}
\hat {\bar G}_{c}^\lambda (i\epsilon_n) = \frac{1}{N/2} \sum_{\bm k}' \hat{\bar {\bm G}}^{\lambda\lambda}_{c,\bm k}(i\epsilon_n)
\end{align}
where $N$ is the total number of sites.
In the DMFT, the bipartite lattice is mapped onto the two effective impurity systems for A and B sublattice systems.
The cavity Green functions at each sublattice are introduced as 
\begin{align}
\hat{\gd}^\lambda_c (i \epsilon_n) = \left( 
\hat{\bar{G}}^\lambda_c (i \epsilon_n)^{-1} + \hat{\Sigma}^\lambda(i \epsilon_n) 
\right) ^{-1}
.
\label{eq_cavity}
\end{align}
The matrix elements 
are explicitly written as
\begin{equation}	
\hat{\gd}_c^\lambda (i \epsilon_n) = 
\begin{pmatrix}
\gd_{c,\uparrow \uparrow}^\lambda (i \epsilon_n) & \gd_{c,\uparrow \downarrow}^\lambda (i \epsilon_n) \\
\gd_{c,\downarrow \uparrow}^\lambda (i \epsilon_n) & \gd_{c,\downarrow \downarrow}^\lambda (i \epsilon_n)
\end{pmatrix} .
\label{eq:2-4}
\end{equation}		
In numerical simulations, we start from the following initial condition:
$\Sigma_{\uparrow\uparrow}(i\epsilon_n) =\Sigma_{\downarrow\downarrow}(i\epsilon_n) =0$ and $\Sigma_{\uparrow\downarrow}(i\epsilon_n) = \Sigma_{\downarrow\uparrow}(i\epsilon_n) = \Sigma_0$.
We then seek for the self-consistent solutions by performing the DMFT iteration.

\subsection{Extension of CT-QMC}

We can relate the cavity Green function to the local Green function by solving the effective impurity problem at each sublattice.
We use the CT-QMC based on $J$-expansion scheme as the impurity solver \cite{otsuki2007continuous}.
However, it cannot deal directly with the transverse magnetism since the cavity Green function is diagonal with respect to spin in the previous method.
Here we extend the $J$-expansion method to allow for finite off-diagonal components as in eq.~(\ref{eq:2-4}).

In the following we omit the site and sublattice indices for simplicity.
In the CT-QMC, the partition function is expanded with respect to $J$.
For the $k$-th order contribution, we have to evaluate the 
quantities
\begin{equation}	
{P}_c(k) = \left\langle T_{\tau} c_{\sigma_k}^\dagger(\tau_k) c_{\sigma_k'}(\tau_k)  \cdots c^\dagger_{\sigma_1}(\tau_1) c_{\sigma_1'}(\tau_1) \right\rangle_c ,
\label{eq:2-4-1}
\end{equation}		
\begin{equation}	
{P}_f(k) = \left\langle T_{\tau} X_{\sigma_k'\sigma_k}(\tau_k) \cdots X_{\sigma_1'\sigma_1}(\tau_1) \right \rangle_f ,
\label{eq:2-4-2}
\end{equation}		
where the interaction representation is used: $A(\tau_i) = e^{\hd_0 \tau_i} A e^{-\hd_0 \tau_i}$.
The averages $\langle \cdots \rangle_c$ and $\langle \cdots \rangle_f$ are taken without interactions between $f$ and $c$ electrons, and we have omitted the summation symbol for spins.
With transverse magnetization, $P_c(q_k)$ cannot be factorized into each spin component.
We then evaluate the determinant of the $k\times k$ matrix whose elements are given by the Fourier transform of cavity Green functions given by eq.~(\ref{eq:2-4}).
Namely eq.~(\ref{eq:2-4-1}) is rewritten using the $k\times k$ matrix $\hat {\cal G}^{(k)}_c$ as
\begin{align}
&{P}_c(k) = \phi_k \det \hat {\cal G}_c^{(k)}
, \label{eq_part_c}
\\
&\hat {\cal G}^{(k)}_{c,ij} \equiv {\cal G}_{c,\sigma_i\sigma_j} (\tau_i - \tau_j) ,
\end{align}
where $\phi_k$ is a sign factor.
On the other hand, the localized-spin contribution $P_f(q_k)$ defined by eq.~(\ref{eq:2-4-2}) can be calculated in the same way as the previous $J$-expansion CT-QMC, as long as the external field is applied along the quantization axis of localized spin.
We have numerically checked that the negative sign problem does not appear even when the transverse components in $P_c(q_k)$ are finite.

The local self-energy can be evaluated by simulation for the effective impurity.
We first define the $t$-matrix by
\begin{equation}	
\hat{G}_c(i \epsilon_n) = \hat{\gd}_c(i \epsilon_n) + \hat{\gd}_c(i \epsilon_n) \ \hat{t}(i \epsilon_n) \ \hat{\gd}_c(i \epsilon_n) ,
\label{eq:2-8}
\end{equation}		
where the left-hand side is the local Green function at the impurity site.
If the solutions are self-consistent, the relation $\hat {\bar G}_c = \hat G_c$ holds.
Note that the $t$-matrix has also off-diagonal components as
\begin{equation}	
\hat{t} (i \epsilon_n) = 
\begin{pmatrix}
t_{\uparrow \uparrow}(i \epsilon_n) & t_{\uparrow \downarrow}(i \epsilon_n) \\
t_{\downarrow \uparrow}(i \epsilon_n) & t_{\downarrow \downarrow}(i \epsilon_n)
\end{pmatrix} 
. \label{eq:2-9}
\end{equation}		
In the CT-QMC simulation, the $t$-matrix is evaluated by the formula
\begin{equation}	
t_{\sigma\sigma'} (\tau) = 
- \frac1{\beta} \left\langle 
\sum_{i,j=1}^k
[ (\hat{\gd}_c^{(k)})^{-1}  ]_{ji} \delta_{\sigma_j \sigma} \delta_{\sigma_i \sigma'}
\delta ( \tau, \tau_{j} - \tau_{i} )
\right\rangle_{\rm MC} .
\label{eq:2-10}
\end{equation}		
The label ``MC'' means the Monte Carlo average.
This expression can be derived in a manner similar to Ref.~\citen{otsuki2007continuous}.
From eqs.~(\ref{eq_cavity}) and (\ref{eq:2-8}), the local self-energy is related to the $t$-matrix as
\begin{equation}	
\hat{\Sigma} (i \epsilon_n) = \hat{t}(i \epsilon_n) \left[ \hat{1} +\hat{\gd}_c (i \epsilon_n) \hat{t}(i \epsilon_n) \right]^{-1} .
\label{eq:2-11}
\end{equation}		
Thus we obtain the new self-energy by solving the effective impurity.
The calculation is repeated until the cavity Green functions converge.

The expectation values for one-body quantities
are derived from $t$-matrix and local Green functions:
\begin{equation}	
\frac J 2
 \langle X_{\sigma \sigma'} \rangle
 = \lim_{n \rightarrow +\infty}t_{\sigma \sigma'} (i\epsilon_n) 
,
\label{eq:3-0-1p} 
\end{equation}		
\begin{equation}	
\langle c_{\sigma}^\dagger c_{\sigma'} \rangle = 
\bar{G}_{c,\sigma'\sigma}(\tau = -0),
\label{eq:3-0-2p} 
\end{equation}		
where eq.~(\ref{eq:3-0-1p}) is obtained by differentiating a Green function $G_c(\tau, \tau')$ with respect to $\tau$ and $\tau'$ \cite{otsuki2008kondo}.
In this paper, we choose $x$ direction as the transverse 
moment,
 so that Green functions in the imaginary-time domain are always real.

We can now evaluate the transverse 
moment perpendicular to the magnetic field.
The transverse magnetic susceptibilities are then calculated
 by measuring the magnetic moments in the presence of small external fields as explained in \S4.

\section{Magnetization Process in Kondo Lattice Model}
\begin{figure}[t]
\begin{center}
\includegraphics[width=80mm]{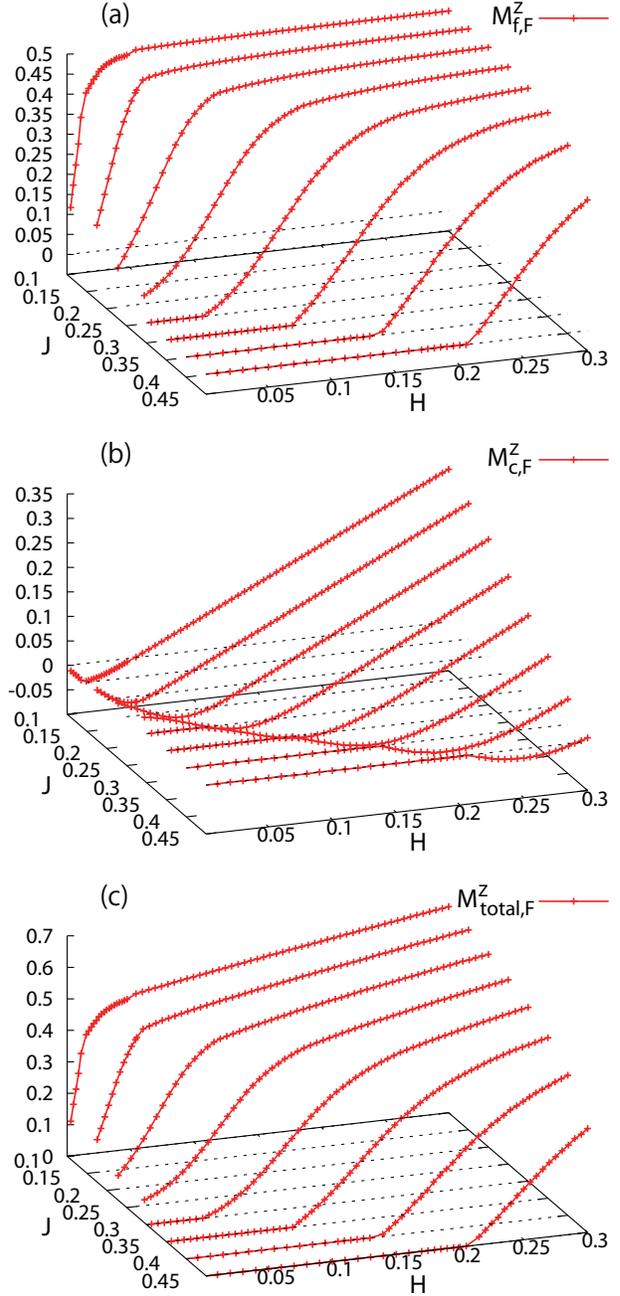}
\caption{(Color online)
Uniform magnetizations of $f$ (a), $c$ (b) and the total electrons (c) in the $J$ - $H_z$ plane at $T=0.01$.
}
\label{fig:F}
\end{center}
\end{figure}

\begin{figure}
\begin{center}
\includegraphics[width=80mm]{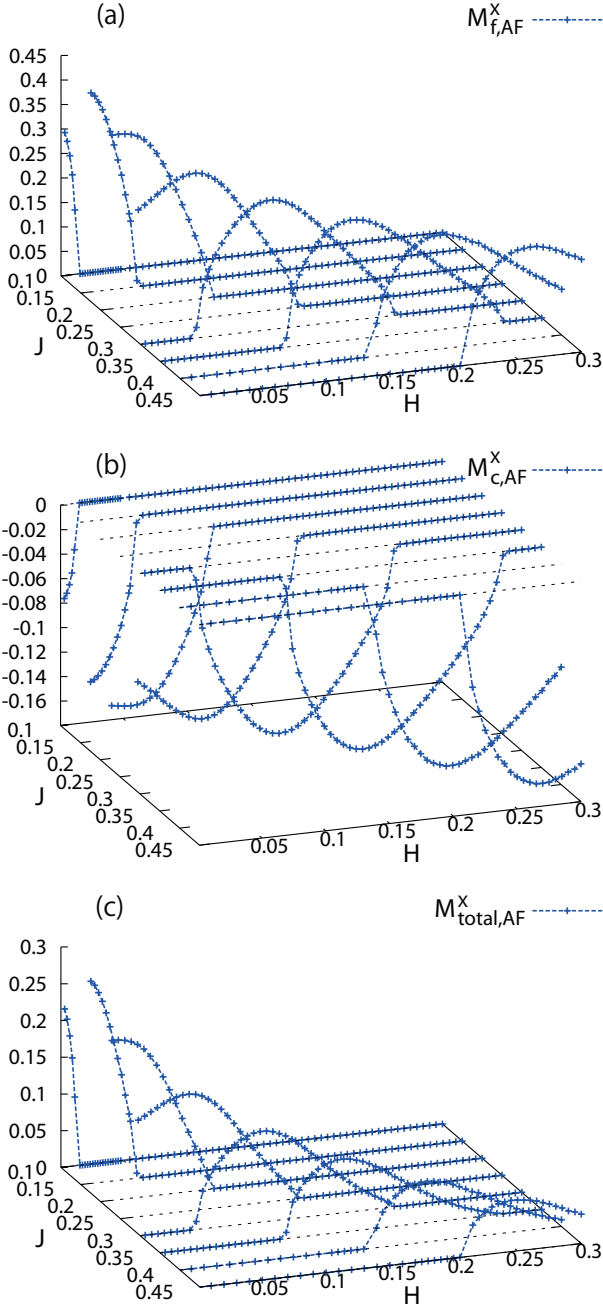}
\caption{(Color online)
Staggered magnetizations of $f$ (a), $c$ (b) and the total electrons (c) in the $J$ - $H_z$ plane at $T=0.01$.
}
\label{fig:AF}
\end{center}
\end{figure}

In this section, we discuss both uniform and staggered magnetizations for several values of interaction $J$ at temperature $T=0.01$.
We define the magnetic moment
\begin{equation}	
M^{\alpha,\lambda}_e = \frac{1}{N/2} \sum_{i\in\{\lambda\}} 
\langle S_{e,i}^{\alpha, \lambda} \rangle
,
\label{eq;4-1} 
\end{equation}		
where $M^{\alpha,\lambda}_e$
is the $\alpha$ ($=x,y,z$)-oriented magnetizations of $e=f,c$
 electron at sublattice $\lambda$.
The uniform (ferro, F) magnetizations $M_{F}^{\alpha}$ and the staggered (antiferro, AF) magnetizations $M_{AF}^\alpha$
 are calculated by $M_{\text{F}}=(M^A + M^B)/2$ and $M_{\text{AF}}=(M^A - M^B)/2$, respectively.
Figures~\ref{fig:F} and \ref{fig:AF} show $M^z_{F}$ and $M^x_{AF}$ in the $J$-$H_z$ plane.
Here (a), (b) and (c) in both figures correspond to the magnetization $M_f$ of $f$ electrons,
 $M_c$ of $c$ electrons and the total value $M_{\text{total}} = M_c + M_f$. 
We have confirmed that the moments $M^x_{F}$ and $M^z_{AF}$ are zero.

As shown in Fig.~\ref{fig:AF}, for $J  <  J_c \simeq 0.27$ with $J_c$ being the quantum critical point,
 the AFM moment is finite at zero magnetic field
 which is consistent with the earlier results. \cite{otsuki2008kondo,hoshino2010itinerant}.
Note that the direction of AFM moments here is perpendicular to the magnetic field.
With increasing field, the staggered magnetizations disappear at sufficiently large $H_z$.
On the other hand, both uniform and staggered $c$ magnetizations take opposite direction to that of $f$ magnetization for weak fields.
The behavior is due to the antiferromagnetic Kondo interaction.

Now we discuss the antiferromagnetism near $J_c$.
The staggered magnetization around $J=0.25$ increases with increasing field at small $H_z$,
which can be seen only near the quantum critical point near $J\lesssim J_c$.
This indicates that the competition between the RKKY interaction and the Kondo effect is responsible for the behavior.
Namely, the magnetic field weakens the Kondo effect because it tends to break the Kondo singlet,
 which results in larger role of the RKKY interaction than the Kondo effect.
Thus the competition between them is reflected in the characteristic magnetic field dependence of magnetizations.
Even larger magnetic field breaks AFM moment as shown in Fig.~\ref{fig:AF}(a--c).

On the other hand, the paramagnetic Kondo insulator is realized for $J > J_c $.
This state is robust against small magnetic fields as shown in Fig.~\ref{fig:F}.
However, transverse magnetizations appear at certain strength of magnetic fields.
The origin for the induced AFM moment is ascribed to weakening the Kondo effect by the magnetic field.
Such behavior has also been reported in earlier study \cite{beach2004field, ohashi2004field}.
It is characteristic in Figs.~\ref{fig:F} and \ref{fig:AF} that
 the AFM moment appears almost simultaneously with appearance of longitudinal uniform magnetization.
With increasing interactions $J$,
 larger magnetic field is necessary for the appearance of the staggered magnetization.
In this case, the magnitude of staggered moments becomes smaller.

Our calculation is qualitatively consistent with the $J$-$H_z$ phase diagram in Ref. \citen{beach2004field}
 for the two-dimensional Kondo lattice model at half filling.

\section{Temperature Dependence of Magnetic Susceptibilities}

\begin{figure}
\begin{center}
\includegraphics[width=50mm]{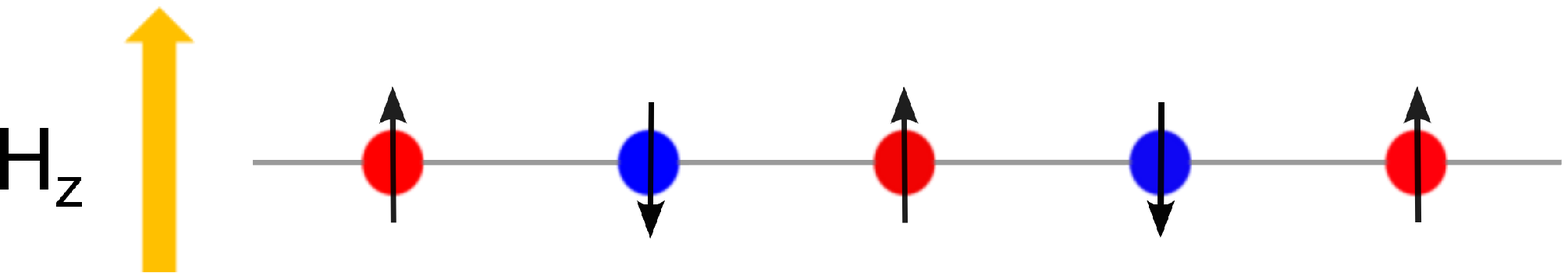}
\includegraphics[width=50mm]{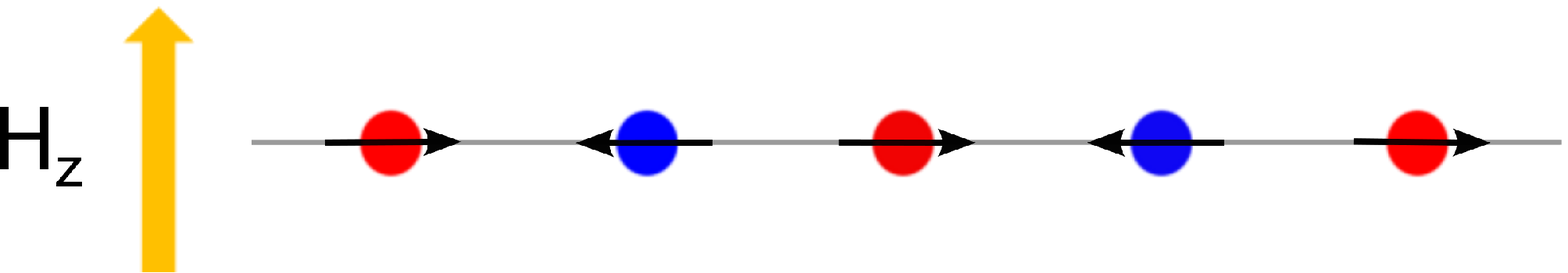}
\caption{(Color online)
Schematic illustrations of situations for calculating longitudinal (upper panel) and transverse (lower panel) susceptibilities.
}
\label{fig:M_H}
\end{center}
\end{figure}

In this section, we discuss temperature dependences of both longitudinal and transverse susceptibilities
 by evaluating the magnetic moment under finite magnetic field.
At zero field,  the antiferromagnetism along $x$ and $z$ axes are degenerate.
With magnetic field along the $z$ axis,
 the longitudinal susceptibility is calculated as
\begin{equation}
\chi_{\text{longi}} = \left.\frac{\delta M^z_{\text{total},{\rm F}}}{H_z} \right| _{\bm M_{\text{AF}}\parallel z} ,
\label{eq:5-1}
\end{equation}
where $\delta M_{\text{total},{\rm F}}^z$ is the change of magnetization under small magnetic field.
On the other hand, the transverse susceptibility is given by
\begin{equation}
\chi_{\text{trans}} = \left.\frac{\delta M^z_{\text{total}, {\rm F}}}{H_z} \right| _{\bm M_{\text{AF}}\parallel x} ,
\label{eq:5-2}
\end{equation}
where the staggered magnetization is along the $x$-axis.
Figure~\ref{fig:M_H} illustrates the situations for eqs.~(\ref{eq:5-1}) and (\ref{eq:5-2}).
When we evaluate the susceptibilities, we must choose a small enough magnetic field.
For this purpose, we take the two different magnetic fields and compute the susceptibilities.
We have confirmed that these two results are almost the same.
The $J$-$H_z$ dependence of magnetizations in \S3 is useful for choosing appropriate magnitude of fields.

\begin{figure}
\begin{center}
\includegraphics[width=80mm]{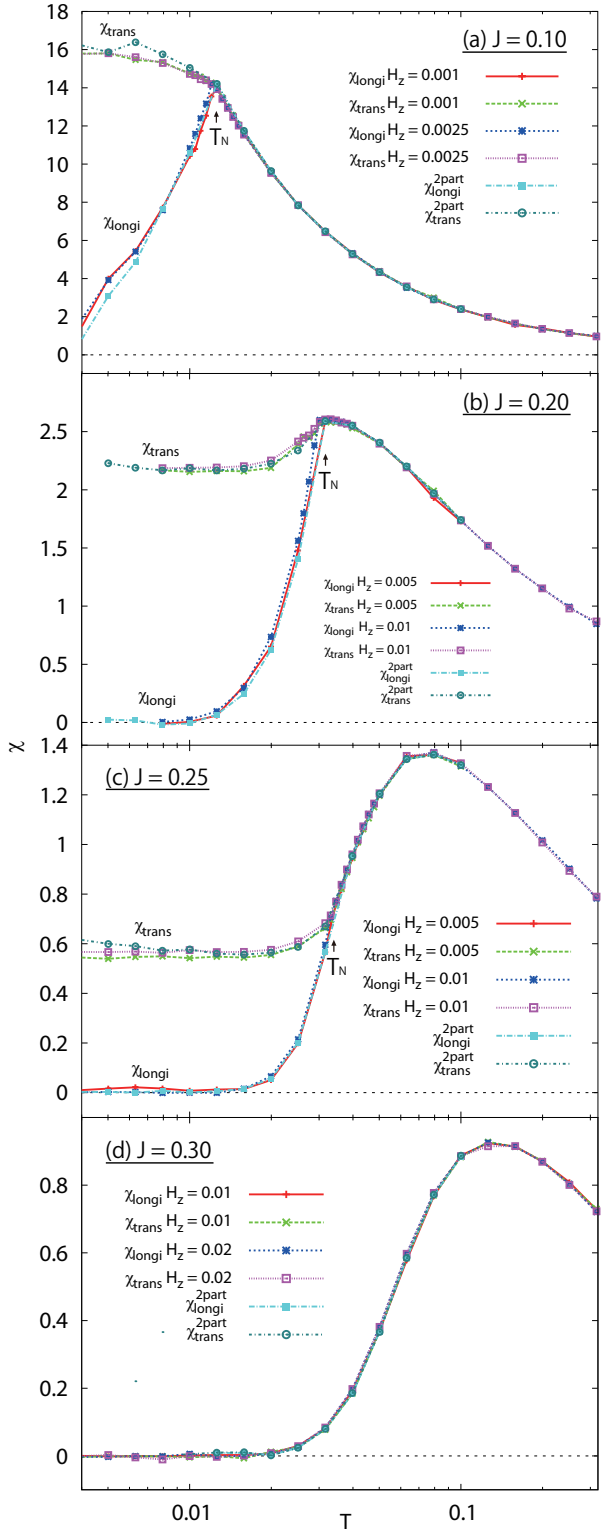}
\caption{(Color online)
Temperature dependence of both longitudinal and transverse susceptibilities in each magnetic field for
 (a) $J=0.10$, (b) $J=0.20$, (c) $J=0.25$ and (d) $J=0.30$.
 The results by two-particle Green functions are indicated as $\chi^{\text{2part}}$.
}
\label{fig:chi_J}
\end{center}
\end{figure}

\begin{figure}
\begin{center}
\includegraphics[width=80mm]{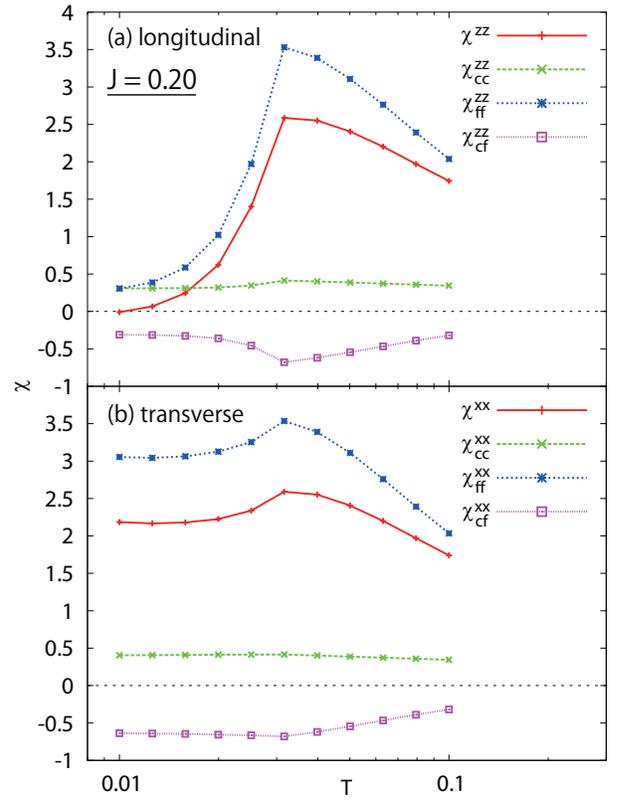}
\caption{(Color online)
Temperature dependence of (a) longitudinal $\chi^{zz}$, and (b) transverse $\chi^{xx}$ susceptibilities for $J=0.20$
calculated by using two-particle Green functions.
The components $\chi_{cc}$, $\chi_{ff}$ and $\chi_{cf}$ are also shown.
}
\label{fig:J20_cf}
\end{center}
\end{figure}

Figure~\ref{fig:chi_J}(a) shows temperature dependence of the susceptibilities for $J=0.10$ where the RKKY interaction is dominant.
Here we have chosen $H_z = 0.0025$ and $ 0.001$ for the magnetic field.
As shown in Fig.~\ref{fig:chi_J}(a), we confirm that the two results with different strength of  magnetic fields are almost the same.
The susceptibilities show the Curie law at high temperatures.
Below the N\'{e}el temperature $T_N$, on the other hand, the longitudinal susceptibility decreases, while the transverse one does not.
This is a typical behavior for small $J$.
It is also characteristic that the transverse susceptibility continue to increase with decreasing temperature even below $T_N$.
This result is unlike the mean field results for the ordinary Heisenberg model where the transverse susceptibility becomes constant below 
the transition temperature.

Next we show the results for $J=0.20$ in Fig.~\ref{fig:chi_J}(b), which is close to the quantum critical point with $J_c  \simeq 0.27$.
We have chosen $H_z = 0.01,\ 0.02$ in this case.
The peak of the longitudinal susceptibility 
 coincides with the N\'{e}el temperature as in the case of $J=0.10$ shown in Fig.~\ref{fig:chi_J}(a).
However, both longitudinal and transverse susceptibilities decrease below the transition temperature for $J=0.20$.
This behavior is due to the Kondo effect, which becomes clearer for larger $J$.

We now take the larger Kondo interaction $J=0.25$ as shown in Fig.~\ref{fig:chi_J}(c).
Similar to the $J=0.20$ case, the longitudinal and transverse susceptibilities decrease below $T_N$ reflecting the Kondo behavior in the AFM phase.
However, the peak temperature of the susceptibility is different from $T_N$.
The peak in the paramagnetic state means the characteristic temperature $T_{\rm KI}$ for the Kondo insulator.
Namely, the collective Kondo singlet state starts to develop below $T_{\rm KI}$.
The peculiar temperature dependence shown in Fig.~\ref{fig:chi_J}(c) is a consequence of the condition $T_N < T_{\rm KI}$.
In the region with $J\lesssim 0.20$, on the other hand, the peak of the susceptibility occurs of $T_N$ because the relation $T_N > T_{\rm KI}$ is satisfied.

Figure~\ref{fig:chi_J}(d) shows temperature dependence of the susceptibilities for $J=0.30$ which is larger than $J_{c}$.
As shown in this figure, there is no AFM order.
Hence we have only $T_{\rm KI}$ as the characteristic energy scale, which gives a peak in the magnetic susceptibility.

\begin{table*}[t]
\begin{center}
\newcolumntype{Y}{>{\centering\arraybackslash}X} 
\begin{tabular}{l|cccc}
\hline
& $J=0.10$ & $J=0.20$ & $J=0.25$ & $J = 0.30$ \\ \hline
\hline
AFM order			& $\bigcirc$ 	& $\bigcirc$ 		& $\bigcirc$		& -- 			\\ \hline
Decrease of $\chi(T)$ below $T_N$ & $z$	& $z$,$x$ 		& $z$,$x$		& --			\\ \hline
Peak position of $\chi(T)$		& $T_N$ 	& $T_N$ 		& $T_{\rm KI}\  (>T_N)$	& $T_{\rm KI}$	\\ \hline
Correspondence to CeT$_2$Al$_{10}$	& --	& CeRu$_2$Al$_{10}$	& CeOs$_2$Al$_{10}$ 	& CeFe$_2$Al$_{10}$		\\ 
\hline
\end{tabular}
\caption{
Proposed correspondence between the Kondo lattice and CeT$_2$Al$_{10}$ family.
Here $z$ and $x$ in the third row represents longitudinal ($z$) and transverse ($x$) components.
}
\label{ta:kekka}
\end{center}
\end{table*}

On the other hand, we also calculate the susceptibility by using the two-particle Green function as described in Appendix.
The results are indicated as $\chi_{\rm longi}^{\rm 2part}$ and $\chi_{\rm trans}^{\rm 2part}$ in Fig.~\ref{fig:chi_J},
 which show a good agreement with those calculated by eqs.~(\ref{eq:5-1}) and (\ref{eq:5-2}) at $T \gtrsim 0.01$.
However, numerical accuracy decreases at lower temperature,
 since the number $(=180)$ of Matsubara frequencies kept in the calculation become too small.

The contributions to the magnetic susceptibility
 can be separated into the parts $\chi_{cc}$, $\chi_{fc}(=\chi_{cf})$ and $\chi_{ff}$.
Firstly, we discuss the uniform susceptibilities $\chi^{\alpha\alpha}_{cc},\ \chi^{\alpha\alpha}_{ff},\ \chi^{\alpha\alpha}_{cf}( = \chi_{fc}^{\alpha\alpha})$, 
$\chi_{\rm longi}^{\rm 2part}$ and $\chi_{\rm trans}^{\rm 2part}$ for $J=0.20$, which are defined by eqs.~(\ref{eq:3-25}--\ref{eq_trans}).
Figures~\ref{fig:J20_cf}(a) and \ref{fig:J20_cf}(b) show the longitudinal ($\alpha = z$) and transverse ($\alpha = x$) susceptibilities, respectively.
The value of $\chi_{cf}^{\alpha\alpha}$ is negative because of the antiferromagnetic $c$-$f$ coupling in the Kondo lattice,
 so that $\chi_{cf}^{\alpha\alpha}$ reduces the total susceptibility $\chi^{\alpha\alpha}$.
The main contribution at $J=0.20$ comes from $\chi_{ff}^{\alpha\alpha}$ and $\chi_{fc}^{\alpha\alpha}$ as shown in the figures.
If we take $J=0.10$, (not shown in the Figure) the dominant contribution comes only from $\chi_{ff}^{\alpha\alpha}$.
At large couplings such as $J=0.30$, all the components equally contribute to the total susceptibility.

Table~\ref{ta:kekka} summarizes the $J$-dependent characteristics of $\chi (T)$.
There is a good correspondence to experimental results of CeT$_2$Al$_{10}$ (T = Ru, Os) \cite{takesaka2010semiconducting,nishioka2009novel,kondo2011high,tanida2010existence}.
Namely, CeRu$_2$Al$_{10}$ shows a behavior similar to the result for $J=0.20$,
 where both longitudinal and transverse susceptibilities decrease below the AFM transition temperature.
The peak position of $\chi (T)$ is nearly the same as $T_N$ in this case.
On the other hand, CeOs$_2$Al$_{10}$ with $T_K$ larger than CeRu$_2$Al$_{10}$ corresponds qualitatively to the result for $J=0.25$.
The susceptibilities also decrease below $T_N$, but the peak position is located in the paramagnetic region with $T>T_N$.
CeFe$_2$Al$_{10}$\cite{takesaka2010semiconducting} is paramagnetic down to experimentally accessible temperature, and hence it corresponds to the result for $J=0.30$.
Thus we roughly understand the temperature dependence of susceptibilities in CeT$_2$Al$_{10}$ family through the isotropic Kondo lattice model.

\section{Summary and discussion}
We have extended the CT-QMC algorithm together with the DMFT so as to deal with transverse magnetizations.
With use of this framework, we have discussed the anisotropic magnetic response inside the AFM phase of the Kondo lattice model.

We have evaluated the temperature dependence of both longitudinal and transverse susceptibilities
 with high accuracy by two ways:
One is to evaluate the magnetic moment under a small field, and the other is to employ the two-particle Green functions.
We have found that the effect of the competition between the RKKY interaction and the Kondo effect is reflected in the decrease of both longitudinal and transverse susceptibilities below the transition temperature, which cannot be explained by the RKKY interaction alone.
The results thus obtained reasonably account for the characteristics in CeT$_2$Al$_{10}$
 : (1) peak positions of temperature dependence of susceptibilities, (2) decrease of susceptibilities below $T_N$ for all the directions.

The present study is still an intermediate step toward understanding the peculiar magnetism of CeT$_2$Al$_{10}$.
In particular we note that the direction of the ordered moment in CeRu$_2$Al$_{10}$
 is different from the easy axis of the susceptibility in the paramagnetic state \cite{tanida2013pressure}.
Consideration of anisotropy in the Kondo lattice is necessary for this aspect, which deserves further study.

\section*{Acknowledgments}
\begin{acknowledgment}
One of the authors (T.K.) was supported by the global COE
program of the Ministry of Education, Culture, Sports,
Science and Technology, Japan (MEXT).

The numerical
calculations were partly performed on supercomputer
in the ISSP, University of Tokyo.

\end{acknowledgment}

\appendix
\section{Susceptibilities Evaluated from Two-Particle Green Functions}
The susceptibilities can be calculated from the two-particle Green functions.
The merit of the method is that we do not need the extrapolation to the zero-field limit.
Furthermore, this method also makes it possible to separate the contributions from $f$ and $c$ electrons.

In the previous study \cite{otsuki2008kondo}, $c$-$c$ and $f$-$f$ correlation functions have been calculated in the framework of the DMFT+CT-QMC.
We show in this Appendix that both $f$-$f$ and $c$-$f$ correlation functions can be derived from the two-particle Green function for $c$ electrons by taking the large frequency limit.

First of all, we consider $c$-$c$ correlation functions.
The wave-vector dependent two-particle Green function is defined by
\begin{align}	
 \chi&_{cc,\bm{kk}'\sigma_1\sigma_2\sigma_3\sigma_4}^{\lambda\lambda'} (\tau_1, \tau_2, \tau_3, \tau_4) \notag  \\
&= \langle T_\tau c_{\bm{k}\lambda\sigma_1}^\dagger (\tau_1)c_{\bm{k}\lambda\sigma_2} (\tau_2)
 c_{\bm{k}'\lambda'\sigma_3}^\dagger (\tau_3)c_{\bm{k}'\lambda'\sigma_4} (\tau_4) \rangle \notag \\
&- \langle T_\tau c_{\bm{k}\lambda\sigma_1}^\dagger (\tau_1)c_{\bm{k}\lambda\sigma_2} (\tau_2) \rangle
 \langle T_\tau c_{\bm{k}'\lambda'\sigma_3}^\dagger (\tau_3)c_{\bm{k}'\lambda'\sigma_4} (\tau_4) \rangle, 
\label{eq:3-1}
\end{align}		
where $c_{\bm{k}\lambda\sigma_i}(\tau_i) = e^{\hd \tau_i}c_{\bm{k}\lambda\sigma_i} e^{-\hd \tau_i} $ is 
the Heisenberg picture,
 and is different from those in eq.~(\ref{eq:2-4-1}).
The Fourier transform of eq.~(\ref{eq:3-1}) is defined by
\begin{equation}	
 \begin{split}
\chi_{cc,\bm{kk}'\sigma_1\sigma_2\sigma_3\sigma_4}^{\lambda\lambda'}(i\epsilon_n, i\epsilon_{n'};i\nu_m)
= \frac1{\beta^2} \int_0^\beta d \tau_1 \cdots \int_0^\beta d \tau_4  \\
\times \chi_{cc,\bm{kk}'\sigma_1\sigma_2\sigma_3\sigma_4}^{\lambda\lambda'} 
(\tau_1, \tau_2, \tau_3, \tau_4 ) e^{i\epsilon_n (\tau_2 - \tau_1)} e^{ i\epsilon_{n'} (\tau_4 - \tau_3)} e^{ i\nu_m (\tau_2 - \tau_3) } ,
\label{eq:3-2}
 \end{split}
\end{equation}		
and its inverse transform by
\begin{align}	
&\chi_{cc,\bm{kk}'\sigma_1\sigma_2\sigma_3\sigma_4}^{\lambda\lambda'}(\tau_1, \tau_2, \tau_3, \tau_4 )= \notag \\
&\frac1{\beta^2} \sum_{nn'm} \chi_{cc,\bm{kk}'\sigma_1\sigma_2\sigma_3\sigma_4}^{\lambda\lambda'} (i\epsilon_n, i\epsilon_{n'}; i\nu_m) \notag \\
&\times e^{-i\epsilon_n (\tau_2 - \tau_1)} e^{-i\epsilon_{n'} (\tau_4 - \tau_3)} e^{-i\nu_m (\tau_2 - \tau_3) } ,
\label{eq:3-2_inv}
\end{align}		
where $\nu_m = 2n\pi/\beta $ is a bosonic Matsubara frequency. 
The Bethe-Salpeter equation relates the two-particle Green function to the vertex part which
 is local in the DMFT but can depend on the sublattice index.
The explicit form is given by \cite{georges1996dynamical}
\begin{equation}	
 \begin{split}
\chi_{cc,\bm{kk}'\sigma_1\sigma_2\sigma_3\sigma_4}^{\lambda\lambda'} (i\epsilon_n, i\epsilon_{n'} ; i\nu_m) 
= \chi_{cc,\bm{k}\sigma_1\sigma_2\sigma_3\sigma_4}^{0,\lambda\lambda'} (i\epsilon_n ; i\nu_m)\delta_{nn'}\delta_{\bm{kk}'} \\
+ \sum_{n_1} \sum_{\lambda_1} \sum_{\bm{k}_1}' \sum_{\sigma_1'\sigma_2'\sigma_3'\sigma_4'} 
\chi_{cc,\bm{k}\sigma_1\sigma_2\sigma_2'\sigma_1'}^{0,\lambda\lambda_1} (i\epsilon_n ; i\nu_m) \\
\times \Gamma_{\sigma_1'\sigma_2'\sigma_3'\sigma_4'}^{\lambda_1}(i\epsilon_n, i\epsilon_{n_1} ; i\nu_m) 
\chi_{cc,\bm{k}_1\bm{k}'\sigma_4'\sigma_3'\sigma_3\sigma_4}^{\lambda_1\lambda'} (i\epsilon_{n_1}, i\epsilon_{n'} ; i\nu_m) ,
\label{eq:3-3}
 \end{split}
\end{equation}		
where $\Gamma^{\lambda_1}_{\sigma_1'\sigma_2'\sigma_3'\sigma_4'}(i\epsilon_n, i\epsilon_{n_1} ; i\nu_m)$ is the vertex part,
 which can be calculated in the effective impurity system\cite{otsuki2008kondo}.
The two-particle Green function without the vertex correction is defined by 
\begin{equation}	
\chi_{cc,\bm{k}\sigma_1\sigma_2\sigma_3\sigma_4}^{0,\lambda\lambda'} (i\epsilon_n ; i\nu_m)
 = - \bar{G}_{c,\bm{k}\sigma_4\sigma_1}^{\lambda'\lambda} (i\epsilon_n+i\nu_m)
\bar{G}_{c,\bm{k}\sigma_2\sigma_3}^{\lambda\lambda'} (i\epsilon_n) ,
\label{eq:3-4}
\end{equation}		
where $\bar{G}_{c,\bm{k}\sigma\sigma'}^{\lambda\lambda'} (i\epsilon_n)$ is given in eq \eqref{eq:2-3}.
Taking the summation over the wave vectors $\bm{k},\bm{k}'$ and the Matsubara frequencies $n,n'$ in eq.~(\ref{eq:3-3}),
 we obtain the 
dynamical susceptibility for $c$ electrons as
\begin{align}	
\chi_{cc,\sigma_1\sigma_2\sigma_3\sigma_4}^{\lambda\lambda'} (i\nu_m) 
\equiv
 \frac14 \int_0^\beta d \tau \langle \tilde{n}_{\sigma_1\sigma_2}^\lambda (\tau) \tilde{n}_{\sigma_3\sigma_4}^{\lambda'} \rangle e^{i\nu_m \tau} \notag \\
= \frac1{4\beta} \sum_{nn'}\frac1{N/2} \sum_{\bm{kk}'}' \chi_{cc,\bm{kk}'\sigma_1\sigma_2\sigma_3\sigma_4}^{\lambda\lambda'} (i\epsilon_n, i\epsilon_{n'} ; i\nu_m) ,
\label{eq:3-4-1}
\end{align}		
where $\tilde{n}_{\sigma\sigma'}^\lambda = {n}_{\sigma\sigma'}^\lambda - \langle n_{\sigma\sigma'}^\lambda \rangle$ and
 $n_{\sigma\sigma'}^\lambda = \sqrt{2/N} \sum_{\bm{k}}' c_{\bm{k}\lambda\sigma}^\dagger c_{\bm{k}\lambda\sigma'}$.

According to the earlier study\cite{hoshino2012DrThes}, we can 
extract
the $f$-$f$ correlation function from the two-particle Green function in eq.~(\ref{eq:3-1}).
Here we show that it is also possible to calculate the $c$-$f$ correlation function in a similar manner.
For this purpose, it is convenient to use the following quantity instead of eq. \eqref{eq:3-1}:
\begin{align}	
\tilde \chi&_{cc,\bm{kk}'\sigma_1\sigma_2\sigma_3\sigma_4}^{\lambda\lambda'} (\tau_1, \tau_2, \tau_3, \tau_4)
\notag \\
&= 
\chi_{cc,\bm{kk}'\sigma_1\sigma_2\sigma_3\sigma_4}^{\lambda\lambda'} (\tau_1, \tau_2, \tau_3, \tau_4)
\notag \\
&- \langle T_\tau c_{\bm{k}\lambda\sigma_1}^\dagger (\tau_1)c_{\bm{k}'\lambda'\sigma_4} (\tau_4) \rangle 
\langle T_\tau c_{\bm{k}\lambda\sigma_2} (\tau_2)c_{\bm{k}'\lambda'\sigma_3}^\dagger (\tau_3) \rangle .
\label{eq:3-10}
\end{align}		
The second term of the right hand side is the Fourier transform of eq.~\eqref{eq:3-4}.
We use eq.~\eqref{eq:3-10} instead of eq.~\eqref{eq:3-1} simply for easier derivation of the formula.
Differentiating both sides of eq.~(\ref{eq:3-10}) with respect to $\tau_3$ and $\tau_4$, we obtain
\begin{align}	
& \frac{\p^2}{ \p \tau_4  \p \tau_3} \tilde \chi_{cc,\bm{kk}'\sigma_1\sigma_2\sigma_3\sigma_4}^{\lambda\lambda'} (\tau_1, \tau_2, \tau_3, \tau_4) \notag \\
&= \langle T_\tau 
c_{\bm{k}\lambda\sigma_1}^\dagger(\tau_1) c_{\bm{k}\lambda\sigma_2}(\tau_2) 
j_{\bm{k}'\lambda'\sigma_3}^+(\tau_3) j_{\bm{k}'\lambda'\sigma_4}(\tau_4)\rangle \notag \\
&- \langle T_\tau 
c_{\bm{k}\lambda\sigma_1}^\dagger(\tau_1) c_{\bm{k}\lambda\sigma_2}(\tau_2) \rangle \langle 
T_\tau j_{\bm{k}'\lambda'\sigma_3}^+(\tau_3) j_{\bm{k}'\lambda'\sigma_4}(\tau_4)\rangle \notag \\
&- \langle T_\tau 
c_{\bm{k}\lambda\sigma_1}^\dagger(\tau_1) j_{\bm{k}'\lambda'\sigma_4}(\tau_4) \rangle \langle 
T_\tau c_{\bm{k}\lambda\sigma_2}(\tau_2) j_{\bm{k}'\lambda'\sigma_3}^+(\tau_3) \rangle \notag \\
&- \delta(\tau_4 - \tau_3) \left( \langle T_\tau 
c_{\bm{k}\lambda\sigma_1}^\dagger(\tau_1) c_{\bm{k}\lambda\sigma_2}(\tau_2) 
\{ j_{\bm{k}'\lambda'\sigma_3}^+, c_{\bm{k}'\lambda'\sigma_4} \}(\tau_4) \rangle \right. \notag \\
&- \left. \langle T_\tau 
c_{\bm{k}\lambda\sigma_1}^\dagger(\tau_1) c_{\bm{k}\lambda\sigma_2}(\tau_2) \rangle \langle 
\{ j_{\bm{k}'\lambda'\sigma_3}^+ , c_{\bm{k}'\lambda'\sigma_4} \} \rangle \right) . 
\label{eq:3-11}
\end{align}		
where $j_{\bm{k}\lambda\sigma}$ and $j_{\bm{k}\lambda\sigma}^+$ are defined by 
\begin{align}	
j_{\bm{k}\lambda\sigma} &= \left. \frac{\p c_{\bm{k}\lambda\sigma}(\tau)}{\p \tau} \right|_{\tau=0} = [\hd, c_{\bm{k}\lambda\sigma}], \label{eq:3-6-1} \\
j_{\bm{k}\lambda\sigma}^+ &= \left. \frac{\p c^\dagger_{\bm{k}\lambda\sigma}(\tau)}{\p \tau} \right|_{\tau=0} = [\hd, c^\dagger_{\bm{k}\lambda\sigma} ] 
=-j_{\bm{k}\lambda\sigma}^\dagger. \label{eq:3-6}
\end{align}		
On the other hand, 
the left hand side of eq.~(\ref{eq:3-11})
 can be expressed by 
 using eq. \eqref{eq:3-2_inv} as
\begin{align}	
& \frac{\p^2}{ \p \tau_4  \p \tau_3} 
\tilde \chi_{cc,\bm{kk}'\sigma_1\sigma_2\sigma_3\sigma_4}^{\lambda\lambda'} (\tau_1, \tau_2, \tau_3, \tau_4) \notag \\
&= \frac{1}{\beta^2} \sum_{nn'm} \epsilon_{n'}(\epsilon_{n'} + \nu_m) 
\tilde \chi_{cc,\bm{kk}'\sigma_1\sigma_2\sigma_3\sigma_4}^{\lambda\lambda'} (i\epsilon_n, i\epsilon_{n'} ; i\nu_m) \notag \\
&\ \ \ \ \ \ \ \ \ \ \times e^{-i\epsilon_n (\tau_2 - \tau_1)} e^{ -i\epsilon_{n'} (\tau_4 - \tau_3)} e^{ -i\nu_m (\tau_2 - \tau_3) } .
\label{eq:3-12}
\end{align}		
With the aid of eq.~\eqref{eq:3-2}, we obtain from eq.~\eqref{eq:3-12}
\begin{align}	
&\epsilon_{n'}(\epsilon_{n'} + \nu_m) 
\tilde \chi_{cc,\bm{kk}'\sigma_1\sigma_2\sigma_3\sigma_4}^{\lambda\lambda'} (i\epsilon_n, i\epsilon_{n'} ; i\nu_m) \notag \\
&= \frac1{\beta^2} \int_0^\beta d \tau_1 d \tau_2 d \tau_3 d \tau_4
\frac{\p^2}{ \p \tau_4  \p \tau_3} 
\tilde \chi_{cc,\bm{kk}'\sigma_1\sigma_2\sigma_3\sigma_4}^{\lambda\lambda'} (\tau_1, \tau_2, \tau_3, \tau_4) \notag \\
& \ \ \ \ \ \ \ \ \times e^{i\epsilon_n (\tau_2 - \tau_1)} e^{ i\epsilon_{n'} (\tau_4 - \tau_3)} e^{ i\nu_m (\tau_2 - \tau_3) } .
\label{eq:3-13}
\end{align}		
We take the limit $n' \rightarrow +\infty$ in eq.~\eqref{eq:3-13} with $m$ fixed.
Then only the terms with $\delta(\tau_4 - \tau_3)$ remain finite.
Taking also the summation over $n$, we arrive at the formula
\begin{align}	
&\sum_{n} \lim_{n' \rightarrow +\infty} \epsilon_{n'}^2 
\tilde \chi_{cc,\bm{kk}'\sigma_1\sigma_2\sigma_3\sigma_4}^{\lambda\lambda'} (i\epsilon_n, i\epsilon_{n'} ; i\nu_m) \notag \\
&= - \int_0^\beta d \tau \left( \langle T_\tau 
c_{\bm{k}\lambda\sigma_1}^\dagger (\tau) c_{\bm{k}\lambda\sigma_2}(\tau) 
\{ j_{\bm{k}'\lambda'\sigma_3}^+, c_{\bm{k}'\lambda'\sigma_4} \} \rangle \right. \notag \\
&\ \ \ \ \ \ \ \ \ \ \ \ \ \ \ \ - \left. \langle 
c_{\bm{k}\lambda\sigma_1}^\dagger c_{\bm{k}\lambda\sigma_2} \rangle \langle 
\{ j_{\bm{k}'\lambda'\sigma_3}^+ , c_{\bm{k}'\lambda'\sigma_4} \} \rangle \right) e^{ i\nu_m \tau} .
\label{eq:3-14}
\end{align}		
The anticommutation relations in eq.~(\ref{eq:3-14}) are then calculated in the Kondo lattice as
\begin{equation}	
 \begin{split}
 \{ c_{\bm{k}\lambda\sigma_1}^\dagger   , j_{\bm{k}\lambda\sigma_2} \} 
= -\left( \mu + \frac{J}2 \right) \delta_{\sigma_1\sigma_2} 
+{J} &\frac1{N/2} \sum_{i\in \{ \lambda \} } X^\lambda_{i\sigma_1\sigma_2}  \\
 \{ j_{\bm{k}'\lambda'\sigma_3}^+ , c_{\bm{k}'\lambda'\sigma_4} \}
=  \left( \mu + \frac{J}2 \right) \delta_{\sigma_3\sigma_4} 
-{J} &\frac1{N/2} \sum_{i'\in \{ \lambda' \} } X^{\lambda'}_{i'\sigma_3\sigma_4} .
\label{eq:3-8}
 \end{split}
\end{equation}		
Substituting eq.~(\ref{eq:3-8}) into eq.~(\ref{eq:3-14}) and taking the summation over the wave vectors $\bm{k}$,$\bm{k}'$,
 we obtain the $c$-$f$ component of the local dynamical susceptibility as
\begin{equation}	
 \begin{split}
\chi_{cf, \sigma_1\sigma_2\sigma_3\sigma_4}^{\lambda\lambda'}(i\nu_m) 
\equiv
 \frac14 \int_0^\beta d \tau 
 \langle T_\tau \tilde{n}_{\sigma_1\sigma_2}^\lambda (\tau) 
 \tilde{X}_{\sigma_3\sigma_4}^{\lambda'} \rangle e^{ i\nu_m \tau} \\
 = \frac1{2J} \frac1{N/2} \sum_{\bm{kk}'}' \sum_{n} \lim_{n' \rightarrow -\infty} \epsilon_{n'}^2 
\tilde \chi_{cc,\bm{kk}'\sigma_1\sigma_2\sigma_3\sigma_4}^{\lambda\lambda'} (i\epsilon_n, i\epsilon_{n'} ; i\nu_m) .
\label{eq:3-15}
 \end{split}
\end{equation}		
where we have defined $\tilde{X}_{\sigma\sigma'}^\lambda = X_{\sigma\sigma'}^\lambda - \langle X_{\sigma\sigma'}^\lambda \rangle$ and
 $X_{\sigma\sigma'}^\lambda = \sqrt{2/N} \sum_{i\in \{ \lambda \} }X_{i\sigma\sigma'}^\lambda $.

Similarly, we can also derive $f$-$c$ component.
In this case, we take the derivative of eq.~\eqref{eq:3-10} with respect to $\tau_1$ and $\tau_2$, and obtain
\begin{equation}	
 \begin{split}
\chi_{fc, \sigma_1\sigma_2\sigma_3\sigma_4}^{\lambda\lambda'}(i\nu_m) 
\equiv
 \frac14 \int_0^\beta d \tau 
 \langle T_\tau \tilde{X}_{\sigma_1\sigma_2}^\lambda (\tau) 
 \tilde{n}_{\sigma_3\sigma_4}^{\lambda'} \rangle e^{ i\nu_m \tau} \\
= \frac1{2J} \frac1{N/2} \sum_{\bm{kk}'}' \sum_{n'} \lim_{n \rightarrow +\infty} \epsilon_n^2 
\tilde \chi_{cc,\bm{kk}'\sigma_1\sigma_2\sigma_3\sigma_4}^{\lambda\lambda'} (i\epsilon_n, i\epsilon_{n'} ; i\nu_m) .
\label{eq:3-20}
 \end{split}
\end{equation}		
Furthermore, if we differentiate eq.~\eqref{eq:3-10} with respect to $\tau_1, \tau_2, \tau_3, \tau_4$, we derive the $f$-$f$ component as \cite{hoshino2012DrThes}
\begin{align}	
 \chi_{ff, \sigma_1\sigma_2\sigma_3\sigma_4}^{\lambda\lambda'}(i\nu_m) 
&= \frac14 
\int_0^\beta d \tau \langle \tilde{X}_{\sigma_1\sigma_2}^\lambda (\tau) \tilde{X}_{\sigma_1\sigma_2}^{\lambda'} \rangle e^{i\nu_m \tau}  \notag \\ 
= \frac{\beta}{4J^2}& \frac1{N/2} \sum_{\bm{kk}'}' \lim_{n\rightarrow +\infty}\lim_{n'\rightarrow -\infty} \notag \\
\times \epsilon_{n}^2  \epsilon_{n'}^2 &
\tilde \chi_{cc,\bm{kk}'\sigma_1\sigma_2\sigma_3\sigma_4}^{\lambda\lambda'} (i\epsilon_n, i\epsilon_{n'} ; i\nu_m) ,
\label{eq:3-9}
\end{align}		
Thus we derive $\chi_{cf}$, $\chi_{fc}$ and $\chi_{ff}$ by taking the high-frequency limit of the two-particle Green function \eqref{eq:3-3} for $c$ electrons.

We now consider the static and uniform components of eqs.~\eqref{eq:3-4-1} and (\ref{eq:3-15}--\ref{eq:3-9}) which are given by
\begin{equation}	
\chi_{ee', \sigma_1\sigma_2\sigma_3\sigma_4} = \frac 1 2 \sum_{\lambda\lambda'} \chi_{ee', \sigma_1\sigma_2\sigma_3\sigma_4}^{\lambda \lambda'}(i\nu_m = 0) .
\label{eq:3-21}
\end{equation}		
The label $e$ or $e'$ corresponds to $c$ or $f$.
We define the uniform susceptibility by
\begin{align}
\chi_{ee'}^{\alpha\alpha'} = 
\frac 1 2 \sum_{\lambda\lambda'}
\int_0^\beta d\tau \langle 
\tilde S_{e}^{\alpha, \lambda}(\tau) \tilde S_{e'}^{\alpha',\lambda'}
\rangle
, \label{eq_suscep}
\end{align}
where $\tilde S_{e}^{\alpha, \lambda} = S_{e}^{\alpha, \lambda} - \langle S_{e}^{\alpha, \lambda} \rangle$ and
 $S_{e}^{\alpha, \lambda} = \sqrt{2/N} \sum_{i\in \{ \lambda \} } S_{i, e}^{\alpha, \lambda} $ with $\alpha,\alpha' = x,y,z$.
If the AFM moments point to the $z$ direction, the longitudinal and transverse susceptibilities are calculated as 
\begin{align}
\chi^{zz}_{ee'} 
= \sum_\sigma \left( \chi_{ee', \sigma\sigma\sigma\sigma} - \chi_{ee', \sigma\sigma\bar{\sigma}\bar{\sigma}} \right) ,
\label{eq:3-25} 
\\
\chi^{xx}_{ee'} 
= \sum_\sigma \left( \chi_{ee', \sigma\bar{\sigma}\bar{\sigma}\sigma} + \chi_{ee', \sigma\bar{\sigma}\sigma\bar{\sigma}} \right) .
\label{eq:3-26}
\end{align}
Finally the total magnetic susceptibility is given by
\begin{align}
\chi_{\rm longi}^{\rm 2part} &= \sum_{ee'} \chi^{zz}_{ee'}
, \label{eq_longi}
\\
\chi_{\rm trans}^{\rm 2part} &= \sum_{ee'} \chi^{xx}_{ee'}
. \label{eq_trans}
\end{align}

Section 4 provides the numerical results for the susceptibilities 
derived from eqs.~(\ref{eq:3-25}--\ref{eq_trans}).

\bibliography{ver3_2014_04_03}

\label{lastpage}
\clearpage

\end{document}